\documentclass[reviewcopy]{elsarticle}
\usepackage{natbib,longtable}
\usepackage{multirow,multicol}
\usepackage{rotating}
\usepackage{graphicx,array}
\usepackage[reviewcopy]{adndt}
\usepackage{amsmath}
\usepackage{amssymb}
\usepackage{lscape}
\usepackage{color}
\begin{document}
\begin{frontmatter}

  \title{Low-energy collisions between electrons and BeH$^+$: \\
  cross sections and rate coefficients for all the vibrational states of the ion}

\author[1,2]{S. Niyonzima}
\author[1,3]{S. Ilie}
\author[1,3]{N. Pop}
\author[1,4,5,6]{J. Zs. Mezei}
\author[7]{K. Chakrabarti}
\author[8]{V. Morel}
\author[8]{B. Peres}
\author[9]{D. A. Little}
\author[4]{K. Hassouni}
\author[10]{\AA. Larson}
\author[11]{A. E. Orel}
\author[5]{D. Benredjem}
\author[8]{A. Bultel}
\author[9]{J. Tennyson}
\author[12]{D. Reiter}
\author[1,5]{I. F. Schneider\corref{cor1}}
\ead{ioan.schneider@univ-lehavre.fr}

\cortext[cor1]{Corresponding author.}

\address[1]{Laboratoire Ondes et Milieux Complexes CNRS$-$Universit{\'{e}} du Havre$-$    Normandie Universit{\'{e}}, 76058 Le Havre, France}
\address[2]{D\'epartement de Physique, Facult\'e des Sciences, Universit\'e du Burundi, B.P. 2700 Bujumbura, Burundi}
\address[3]{Fundamental of Physics for Engineers Department, Politehnica University Timisoara,  300223 Timisoara, Romania}
\address[4]{Laboratoire des Sciences des Proc\'ed\'es et des Mat\'eriaux, CNRS$-$Universit\'e Paris 13$-$USPC, 93430 Villetaneuse, France}
\address[5]{Laboratoire Aim\'{e}-Cotton, CNRS$-$Universit\'e Paris-Sud$-$ENS Cachan$-$Universit\'e Paris-Saclay, 91405 Orsay, France}
\address[6]{Institute of Nuclear Research, Hungarian Academy of Sciences, Debrecen, Hungary}
\address[7]{Department of Mathematics, Scottish Church College, Calcutta 700 006, India}
\address[8]{CORIA CNRS$-$Universit\'{e} de Rouen$-$Universit{\'{e}} Normandie, F-76801 Saint-Etienne du Rouvray, France}
\address[9]{Department of Physics and Astronomy, University College London, WC1E 6BT London, UK}
\address[10]{Department of Physics, Stockholm University, AlbaNova University Center, S-106 91 Stockholm, Sweden}
\address[11]{Department of Chemical Engineering and Materials Science, University of California, Davis, California 95616, USA}
\address[12]{Institute of Energy and Climate Research-Plasma Physics, Forschungszentrum J\"ulich GmbH Association EURATOM-FZJ, Partner in Trilateral Cluster, 52425 J\"ulich, Germany}

\begin{abstract}
We provide cross sections and Maxwell rate coefficients for reactive collisions of slow electrons with BeH$^+$ ions on all the eighteen vibrational levels ($X{^{1}\Sigma^{+}},v_{i}^{+}=0,1,2,\dots,17$) using a Multichannel Quantum Defect Theory (MQDT) - type approach. These data on dissociative recombination, vibrational excitation and vibrational de-excitation are relevant for magnetic confinement fusion edge plasma modelling and spectroscopy, in devices with beryllium based main chamber materials, such as the International Thermonuclear Experimental Reactor (ITER) and the Joint European Torus (JET). Our results are presented in graphical form and as fitted analytical functions, the parameters of which are organized in tables.
\end{abstract}


\end{frontmatter}

\clearpage

\tableofcontents
\listofDtables
\listofDfigures

\clearpage

\section{Introduction}

Even though various isotopes of light elements can be coupled to achieve
thermonuclear fusion energy release, the next generation of thermonuclear fusion
reactors will use deuterium-tritium (D-T) reactions, by far most efficient and
accessible plasma fuel for fusion reactors and power plants. As discussed in
detail elsewhere \cite{federici2001,Pamela2007,Reiter2012,Anderl1999,shimada2007,Brooks1997}, beryllium (Be) is meant to enter the composition of the wall of the future fusion devices (ITER).
Its performance on preventing tritium retention and, meanwhile, still
keeping the benefits of a low Z material (low fuel dilution), is currently being
tested in the JET \cite{Reiter2012,Anderl1999}.
According to current plans, tungsten (W) will be the plasma facing material in
the high heat flux components (the entire divertor). These materials (Be and W)
are expected to provide sufficiently low fuel retention, plasma impurity levels,
neutron damage, and sufficient heat removal capabilities in the divertor and
then, meet ITER requirements \cite{Pamela2007}. 
The key challenge in the use of beryllium as main chamber material
for experimental and commercial fusion devices is to understand, predict and
control the characteristics of the thermonuclear burning plasma, the plasma edge
regimes that result in acceptable erosion performance and the divertor plasma
(heat and particle exhaust, impurity control, lifetime). 

Due the the low mass ratio between beryllium and the D, T plasma fuel ions, beryllium erodes rather easily under plasma exposure by physical and chemical sputtering, a process which releases Be, Be$^+$, and other impurities into the plasma. Significant fractions of the eroded beryllium will be transported towards the divertor and will form compounds with the fuel atoms, molecules and/or molecular ions. Therefore, BeH as well as BeD and BeT molecules are expected to appear in a significant (spectroscopically detectable) amount in the edge and divertor plasmas. Various source mechanisms may lead to their formation, either surface or volumetric (particle rearrangement) processes. The particle interchange reaction \cite{Nishijima2008},
\begin{equation}\label{eq1}
\mathrm{Be}^{+} + \mathrm{X}_{2} \longrightarrow \mathrm{BeX}+\mathrm{X}^{+}, 
\end{equation}
where X denotes one of the fuel atoms H, D or T, was suggested as one, possibly dominant, volumetric BeX formation channel, when X$_2$ is in its vibrational ground state. However, the exo- or endothermicity will greatly depend upon the vibrational state of X$_2$ and other channels, including electron transfer channels, may also be open for BeX formation in fusion divertor plasma conditions.
The involved particles - atoms, molecules and molecular ions - follow their often quite complex transport pathways in the edge plasma and take part in the complex reactions determining the plasma composition. Its detailed modeling by taking into account reactions between all present  species is necessary, first of all, for interpretation of molecular and atomic line spectroscopy and also to understand and predict the overall plasma edge dynamics and the divertor region behavior in particular.

In principle, the rate of beryllium erosion in fusion devices can be measured by spectroscopical techniques of all the states of the atoms and molecules, so primarily of Be, Be$^+$, Be$_2^+$, BeX, BeX$^+$, BeX$_2$, BeX$_2^+$. In order to provide a quantitative interpretation of such spectroscopic measurements, one needs a complete set of rate coefficients for excitation, ionization and the various atomic and molecular ions break-up reactions. The inelastic electron-impact processes of vibrationally excited beryllium monohydride BeH$^+$ play a key role in the reaction kinetics of low-temperature plasmas in general, and potentially also in certain cold regions of fusion reactor relevant (e.g. the divertor) plasmas. In order to model and diagnose plasmas containing BeH$^+$ it is essential to build a complete database of cross-sections and rate coefficients for electron-impact collision processes. Knowing the loss rates of BeH$^+$ 
(and isotopologues) as well as absolutely calibrated spectroscopic emission from this molecular ion will allow to draw conclusions also about the BeH$^+$ formation rates.

The BeH$^{+}$ ion is subject to Dissociative Recombination (DR), competed by Vibrational Transitions (VT) - excitation/de-excitation (VE/VdE) - and 
Dissociative Excitation (DE) respectively \cite{dr2013}:
\begin{equation}\label{eq2}
 \mathrm{BeH}^{+}(v_{i}^{+}) +\mathrm{e^{-}} \longrightarrow  \mathrm{Be + H},\quad \mathrm{BeH}^{+}(v_{f}^{+})+\mathrm{e}^{-},\quad\mathrm{Be}^{+}+\mathrm{H}+\mathrm{e}^{-},
\end{equation}
\noindent
$(v_{i}^{+})/(v_{f}^{+})$ standing for the initial$/$final vibrational levels of the cation.

Whereas for numerous ions measurements of these reactive collisions have been performed in magnetostatic or electrostatic storage rings (multipass experiments using merged electron and ion beams) \cite{dr2013}, this is certainly not the case for BeH$^{+}$, beryllium being highly toxic.

In the current study, we performed large scale computations of cross sections for the reactive collisions 
DR, VE and VdE displayed in eq. (\ref{eq2}), as well as of the corresponding rate coefficients.

More specifically, we have used the molecular structure data computed by some of us 
\cite{roos:09} in order to model the dynamics of these reactions by our
stepwise MQDT-method, neglecting the rotational structure and interactions \cite{Giusti:80,IFS2000a,Valcu1998,niyonzima2013}.
Whereas our previous works \cite{roos:09,niyonzima2013} restricted to the ground and three lowest vibrationally-excited levels of BeH$^{+}$, we have extended here our analysis to the whole range of its vibrational states, i.e. up to $v_{i}^{+}=17$.

After briefly reminding the major ideas and steps of our MQDT method, 
including the main features and parameters of the computational part (section \ref{theoretical}), we present the cross sections and rate coefficients below $2.7$ eV - dissociation threshold -
and below $5000$ K respectively (section \ref{res_disc}). The coefficients appearing in the analytical functions used to
fit the rate coefficients are organized in tables.

At
higher energy of the incident electron, a further competing process,
dissociation excitation, will become effective. The collisional data for this range are the subject of ongoing calculations.

\section{Theoretical approach of the dynamics}\label{theoretical}

In this paper, we  use an MQDT-type method to study the electron-impact collision processes:
\begin{equation}\label{eq:indirect}
\mathrm{BeH}^{+}(v_{i}^{+}) +\mathrm{e}^{-}
\longrightarrow \mathrm{BeH}^{*},
\mathrm{BeH}^{**} \longrightarrow  \left\{
			\begin{array}{ll}
			 \mathrm{Be + H}\\
			 \mathrm{BeH}^{+}(v_{f}^{+})+\mathrm{e}^{-}.\\
			  \end{array}
			  \right.
\end{equation}

\noindent
resulting from the quantum interference between the \textit{direct} mechanism - involving the doubly excited resonant states BeH$^{**}$ - and the \textit{indirect} one - occurring via 
Rydberg singly-excited predissociating states BeH$^{*}$.

A detailed description of our theoretical approach was given in \cite{niyonzima2013}.
Its main steps are the following:

\textit{i) Building of the interaction matrix} $\boldsymbol{\mathcal{V}}$:
It is based on the computed \cite{roos:09,niyonzima2013}
Rydberg-valence couplings within a quasi-diabatic representation of molecular states of the neutral system. The matrix elements of this matrix correspond to - and are accordingly indexed with - all the possible pairs of channels. More specifically, for 
a given electronic total angular momentum quantum number $\Lambda$ 
and 
a given symmetry (total electronic spin singlet/triplet) 
of the neutral, our formalism rely on 
\textit{ionization} channels - 
	labelled by 
	the vibrational quantum number of the cation $v^+$ 
	and 
	the orbital quantum number $l$ of the incident/Rydberg electron - 
and on
\textit{dissociation} channels
	labelled $d_j$.

\textit{ii) Computation of the reaction matrix} $\boldsymbol{\mathcal{K}}$: it is performed by adopting for the Lippman-Schwinger integral equation the second-order perturbative solution
\cite{Ngassam2003a,Florescu2003,Motapon2006}, 
written in operatorial form as:
\begin{equation}\label{eq:solveK}
\boldsymbol{\mathcal{K}}= \boldsymbol{\mathcal{V}} + \boldsymbol{\mathcal{V}}{\frac{1}{E-\boldsymbol{H_0}}}\boldsymbol{\mathcal{V}},
\end{equation}
\noindent
$\boldsymbol{H_0}$ being the Hamiltonian of the molecular system under study in which the Rydberg-valence interaction is neglected.  

\textit{iii) Computation of the eigenchannel wavefunctions:} It relies on the eigenvectors  and eigenvalues of the reaction matrix $\boldsymbol{\mathcal{K}}$, i.e. the columns of the matrix $\boldsymbol{{U}}$ and the elements of the diagonal matrix $\boldsymbol{\tan(\eta)}$ respectively : 
\begin{equation}
\boldsymbol{\mathcal{K}U}= -\frac{1}{\pi}\boldsymbol{\tan(\eta)U}
\end{equation}
\noindent
where the non-vanishing elements of the diagonal matrix $\boldsymbol\eta$ are the phaseshifts introduced into the wavefunctions by the short-range interactions.

\textit{iv) Frame transformation from the Born-Oppenheimer representation to the close-coupling one:} It is performed via the matrices 
$\boldsymbol{\mathcal{C}}$ and $\boldsymbol{\mathcal{S}}$, 
built on the basis of the matrices $\boldsymbol{{U}}$ and $\boldsymbol\eta$ and on the  quantum defect characterizing the incident/Rydberg electron, $\mu_{l}^{\Lambda}(R)$.

\textit{v) Construction of the generalized scattering 
matrix $\boldsymbol{\mathcal{X}}$}, eventually split in blocks associated to open and/or closed (o and/or c respectively) channels:

\begin{equation}
\boldsymbol{\mathcal{X}}=\frac{\boldsymbol{\mathcal{C}}+i\boldsymbol{\mathcal{S}}}{\boldsymbol{\mathcal{C}}-i\boldsymbol{\mathcal{S}}}
\qquad
\boldsymbol{\mathcal{X}}= \left(\begin{array}{cc} \boldsymbol{X_{oo}} & \boldsymbol{X_{oc}}\\
                   \boldsymbol{X_{co}} & \boldsymbol{X_{cc}} \end{array} \right).
\end{equation}

\textit{vi) Construction of the physical scattering matrix $\boldsymbol{\mathcal{S}}$}, whose elements link mutually the open channels exclusively, given by
\cite{Seaton1983}:
\begin{equation}\label{eq:solve3}
\boldsymbol{S}=\boldsymbol{X_{oo}}-\boldsymbol{X_{oc}}\frac{1}{X_{cc}-\exp(-i2\pi\boldsymbol{ \nu})}\boldsymbol{X_{co}}.
\end{equation}
\noindent
Here the matrix $\exp(-i2\pi\boldsymbol{ \nu})$ is diagonal and relies on the effective quantum numbers $\nu_{v^{+}}$ associated to the vibrational thresholds of the closed channels.

\textit{vii) Computation of the cross-sections:}
Given the target cation on its level $v_i^+$, its impact with an electron of energy $\varepsilon$ results either in dissociative recombination or in vibrational excitation/de-excitation according to the formulae:

\begin{equation}\label{eqDR}
\sigma _{{\rm diss} \leftarrow v_{i}^{+}}=\sum_{\Lambda,sym} \sigma _{{\rm diss} \leftarrow v_{i}^{+}}^{{\rm sym},\Lambda},
\quad \quad
\sigma _{{\rm diss} \leftarrow v_{i}^{+}}^{{\rm sym},\Lambda}=\frac{\pi}{4\varepsilon} \rho^{{\rm sym},\Lambda}\sum_{l,j}\mid S_{d_{j},lv_{i}^{+}}\mid^2,
\end{equation}

\begin{equation}\label{eqVE_VdE}
\sigma _{v_{f}^{+} \leftarrow v_{i}^{+}}=\sum_{\Lambda,sym} \sigma _{v_{f}^{+}\leftarrow v_{i}^{+}}^{{\rm sym},\Lambda}, 
\quad \quad
\sigma _{v_{f}^{+} \leftarrow v_{i}^{+}}^{{\rm sym},
\Lambda}=\frac{\pi}{4\varepsilon} \rho^{{\rm sym},\Lambda}\sum_{l,l'}\mid S_{l' v_{f}^{+},lv_{i}^{+}}-\delta_{l'l}\delta_{v_{i}^{+}v_{f}^{+}}\mid^2,
\end{equation}

\noindent
where, $\rho^{{\rm sym},\Lambda}$ stands for the ratio between the multiplicity of the involved electronic states of BeH and that of the target, BeH$^+$.

\section{Results}\label{res_disc}

Using the available molecular data - quasi-diabatic potential energy curves and electronic couplings for \ensuremath{^{2}\Pi}, \ensuremath{^{2}\Sigma^{+}} and
\ensuremath{^{2}\Delta} states displayed 
in Figure 1 of \cite{niyonzima2013} (for more details see as well \cite{roos:09,Strömholm1995}) - we have extended our previous calculations
- initially restricted to the ground and weakly excited vibrational states 
- to 
{\textit{all}} vibrational levels (up to $v_i^+ =17$) of the ground electronic state.
The energy of the electron is inferior to $2.7$ eV, this value corresponding to the dissociation threshold of the ground state ion.

In all the panels of graphs
\ref{fig:2}-\ref{fig:7}, the vertical green lines mark the energy or the temperature below
which the calculations are the most accurate. Indeed, above these thresholds,
these calculations neglect the role of the dissociative excitation of the ion. Nevertheless, the
data displayed continue to be reasonably correct above the thresholds because
this process weakly affects the DR and the VT at low energy/temperature.

In order to illustrate the vibrational dependence of the dissociative recombination and to compare the two mechanisms - direct and total - we display in Graphs \ref{fig:2}, \ref{fig:3}, and \ref{fig:4} the corresponding cross sections. 
 One may notice that the total (direct {\textit{coherently}} added to indirect) cross section is characterized by 
resonant captures into Rydberg states, but that they do not contribute too much in average.

Similar features characterize the cross sections of competitive processes, VE and VdE,
but we do not display them here, restricting ourselves to show only their Maxwell rate coefficients.

Indeed,  graphs \ref{fig:5}, \ref{fig:6}, and \ref{fig:7} give the whole ensemble of  rate coefficients available for the state-to-state kinetics of BeH$^+$. They illustrate the dominance of the  DR in its $v_{i}^{+}=0-9$ levels at low electronic temperature, while the VdE becomes more important than the other processes for initial vibrational states $v_{i}^{+}>9$.

Graph
\ref{fig:8} provides the comparison between the DR rate coefficients and the 
{\textit{global}} - i.e. coming from the sum over all the possible final levels - vibrational transitions rate coefficients. One may notice that the excitation process becomes a notable competitor to DR and VdE above $1000$ K only.

In order to allow the versatile implementation of the rate coefficients  shown in 
Graph
\ref{fig:5}-\ref{fig:7} in kinetics modelling codes, we have fitted them with generalized Arrhenius-type formulas:

\begin{equation}\label{eqn:BeH_DR_Interpolation}
k_{(BeH^+,L)}^{P}(T_e) = A_{L} \, T_e^{\alpha_{L}} \, \exp\left[-\sum_{j=1}^{7}\frac{B_{L}(j)}{j\,\,T_e^j}\right],
\end{equation}

\noindent  over the electron temperature range $100$~K~$\leq$~$T_e$~$\leq$~$5000$~K, 'P' and 'L' standing for Processes (DR or VT) and Levels ($v^+_i$ or $v^+_i\to v^+_f$) respectively. 
The coefficients for Dissociative Recombination (when P corresponds to DR) $A_{v^+_i}$, $\alpha_{v^+_i}$ and $B_{v^+_i}(j)$ are displayed in Table \ref{tab:BeH_DR_Interpolation}, and those for Vibrational Transitions (when P corresponds to VT, i.e. VE and VdE)
 $A_{{v^+_i}\to {v^+_f}}$, $\alpha_{{v^+_i}\to {v^+_f}}$ and $B_{{v^+_i}\to {v^+_f}}(j)$ 
are given in Tables \ref{tab:BeH_VE_Interpolation0-1}-\ref{tab:BeH_VE_Interpolation16-17}. The values interpolated using the equation (\ref{eqn:BeH_DR_Interpolation}) agree with the MQDT-computed ones within a few percent, and are represented in 
Graphs
\ref{fig:5}-\ref{fig:7}.

\section{Conclusions}

The present paper  provides a complete state-to-state information of the BeH$^{+}$ reactive collisions with electrons, illustrating quantitatively the competition between the
vibrational transitions and dissociative recombination. We display the cross
sections or/and the Maxwell rate coefficients for the molecular ion in
all of its initial vibrational states and for the entire range of energies of the incident electron below the ion dissociation threshold.

Arrhenius-type formulas are used in order to fit the rate coefficients as function
of $T_{e}$, the electronic temperature. These rate coefficients strongly depend on 
the initial vibrational level of the molecular ion.

These data are addressed to the fusion community - being relevant to the modeling
of the edge of the fusion plasma as well as for divertor conditions - and, more generally, to the modelers of any beryllium-containing-plasmas, produced in laboratory experiments, industrial processing and natural environments. No experimental work concerning the electron$/$BeH$^+$ collisions can be found in the literature.

Further studies, devoted to higher energy and consequently taking into account the 
dissociative excitation \cite{Kalyan2013}, as well as others, extending to isotopomers of BeH$^+$ (BeD$^+$, BeT$^+$) are the object of ongoing work. All these data, as well as the presently displayed ones will be of huge importance for modeling of the  plasma/wall interaction \cite{Reiter2012}.

\ack
We acknowledge the French LabEx EMC$^3$, via the project PicoLIBS (No. ANR-12-BS05-0011-01), the BIOENGINE project (sponsored by the European fund FEDER and the French CPER), the Fédération de Recherche Fusion par Confinement Magnétique - ITER and the European COST Program CM1401 (Our Astrochemical History). AEO acknowledges support from the National Science Foundation, Grant No PHY-11-60611. In addition some of this material is based on work done while AEO was serving at the NSF. {\AA}L acknowledges support from the Swedish Research council, Grant No. 2014-4164. 
NP and SI acknowledge the Sectoral Operational Programme Human Resources Development (SOP HRD), ID134378 financed from the European Social Fund and by the Romanian Government.

\newpage

\GraphExplanation

\section*{Graph \ref{fig:2}-\ref{fig:4}. Dissociative recombination cross sections for all the vibrational levels of BeH$^+$ in its ground electronic state.}
\begin{tabular}{@{}p{2in}p{6in}@{}}
Ordinate		& Cross section in cm$^2$ \\
Abscissa	& Collision (electron) energy in eV\\
Black solid line	& Total (direct and indirect) process\\
Red dashed line	& Direct process\\
Green vertical line	& Precision limit, for details see the text\\
\end{tabular}

\section*{Graph \ref{fig:5}-\ref{fig:7}. Dissociative recombination, vibrational excitation and de-excitation  Maxwell rate coefficients for  all the vibrational levels of BeH$^+$ in its ground electronic state.}
\begin{tabular}{@{}p{2in}p{6in}@{}}
Ordinate		& Maxwell rate coefficient in cm$^3\cdot$s$^{-1}$ \\
Abscissa	& Electron temperature in K\\
Thick black line	& DR Maxwell rate coefficient \\
Thin coloured lines	& Vibrational excitation rate coefficients\\
Coloured lines	with symbols & Vibrational de-excitation rate coefficients\\
Green vertical line	& Precision limit, for details see the text\\
\end{tabular}

\section*{Graph \ref{fig:8}. Global (summed-up for all possible final states) dissociative recombination, vibrational excitation and de-excitation  Maxwell rate coefficients for BeH$^+$ in its ground electronic state.}
\begin{tabular}{@{}p{2in}p{6in}@{}}
Ordinate		& Maxwell rate coefficient in cm$^3\cdot$s$^{-1}$ \\
Abscissa	& Electron temperature in K\\
\end{tabular}

\newpage

 \begin{Dfigures}[h]
\includegraphics[width=0.9\linewidth]{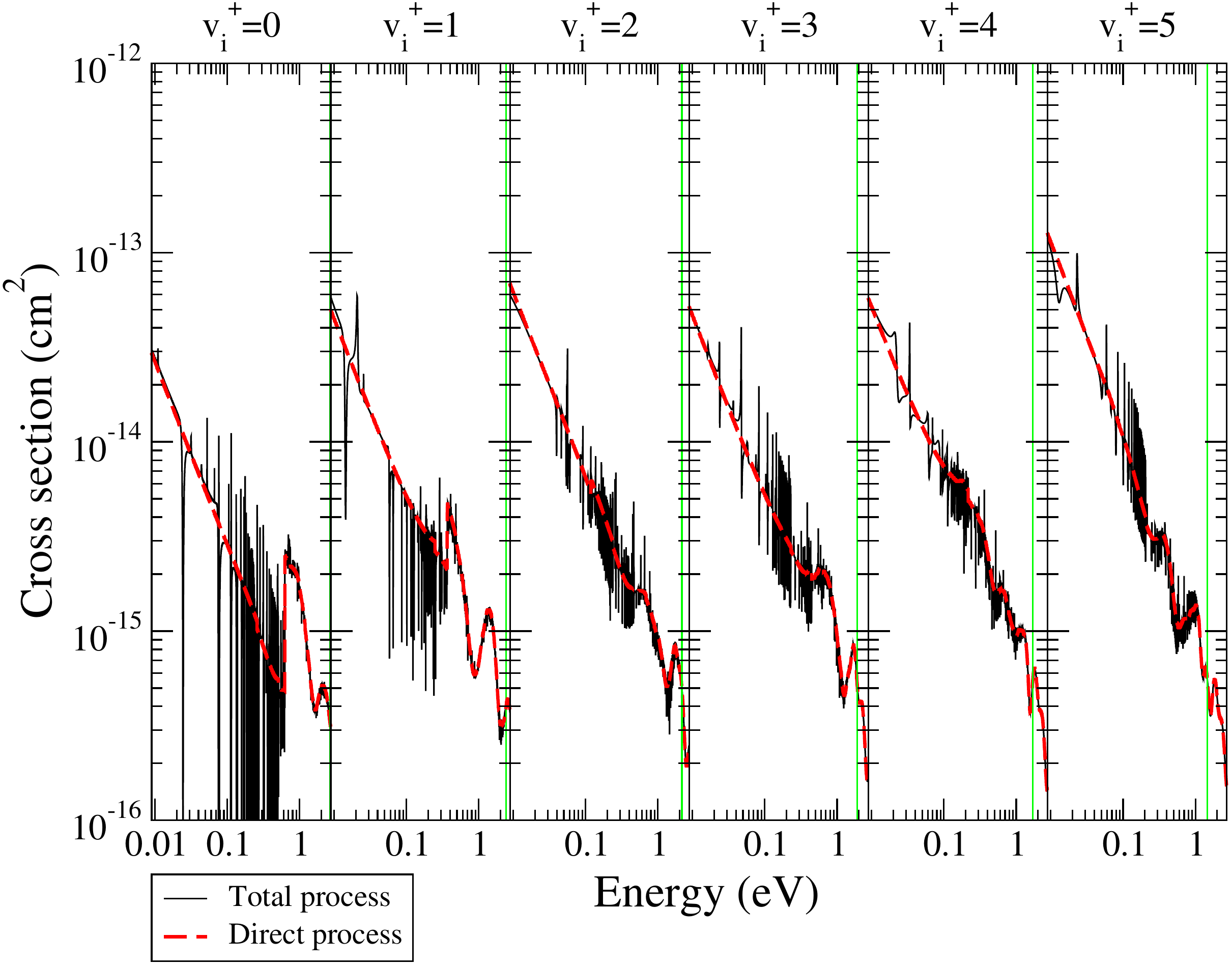} 
\caption{\label{fig:2}  Dissociative recombination cross sections of ground ($v_{i}^{+}=0$) and excited ($v_{i}^{+}=1,...,5$) {\rm BeH$^{+}$} in its electronic ground state. Direct mechanism: dashed thick line, total (direct and indirect) mechanism: continuous thin line.}
\end{Dfigures}

\clearpage

 \begin{Dfigures}[h]
    \includegraphics[width=0.9\linewidth]{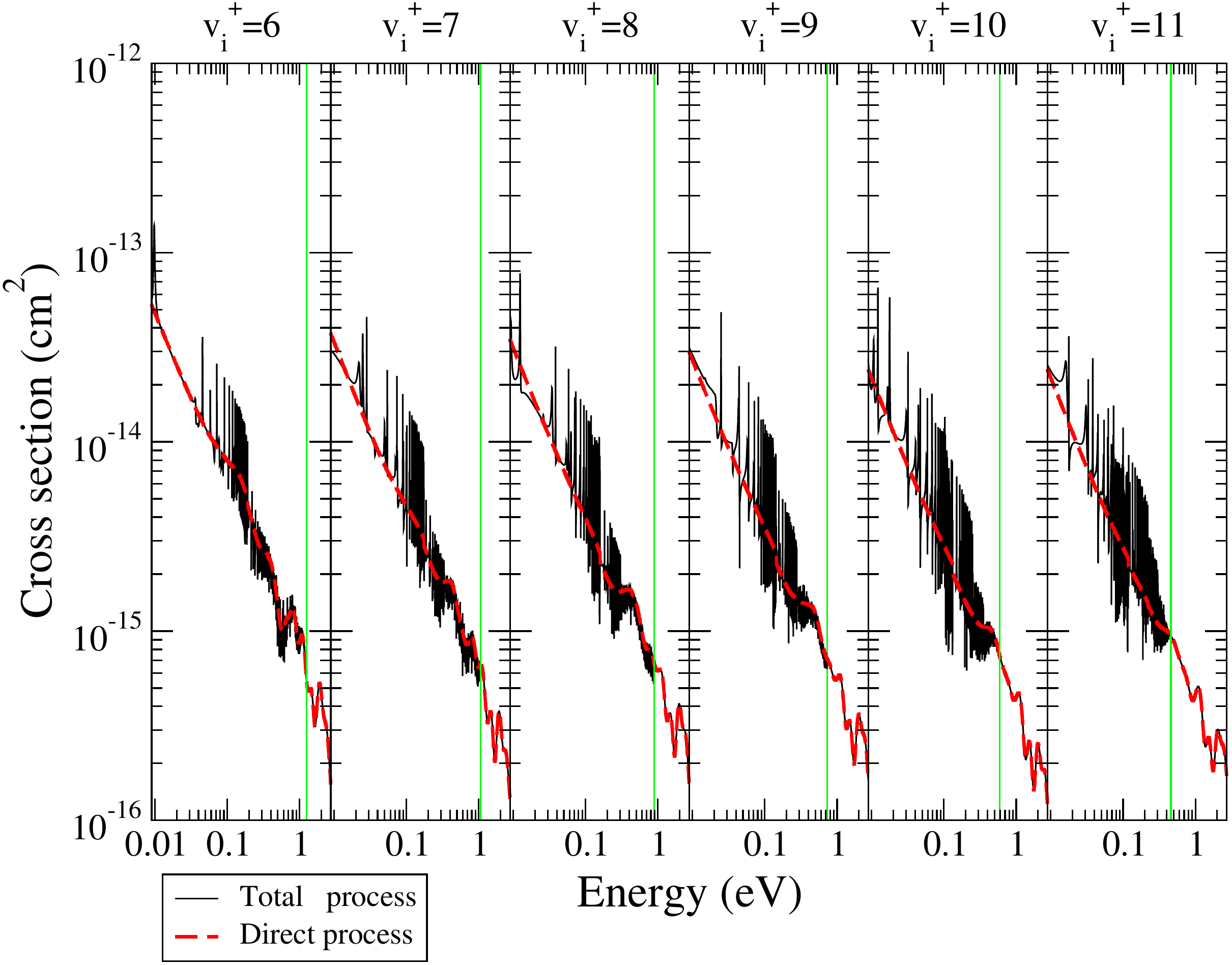}     
     \caption{\label{fig:3} Dissociative recombination cross sections of excited ($v_{i}^{+}=6,7,...,11$) {\rm BeH$^{+}$} in its electronic ground state. Direct mechanism: dashed thick line, total (direct and indirect) mechanism: continuous thin line.}
\end{Dfigures}

\clearpage

\begin{Dfigures}[h]
\includegraphics[width=0.9\linewidth]{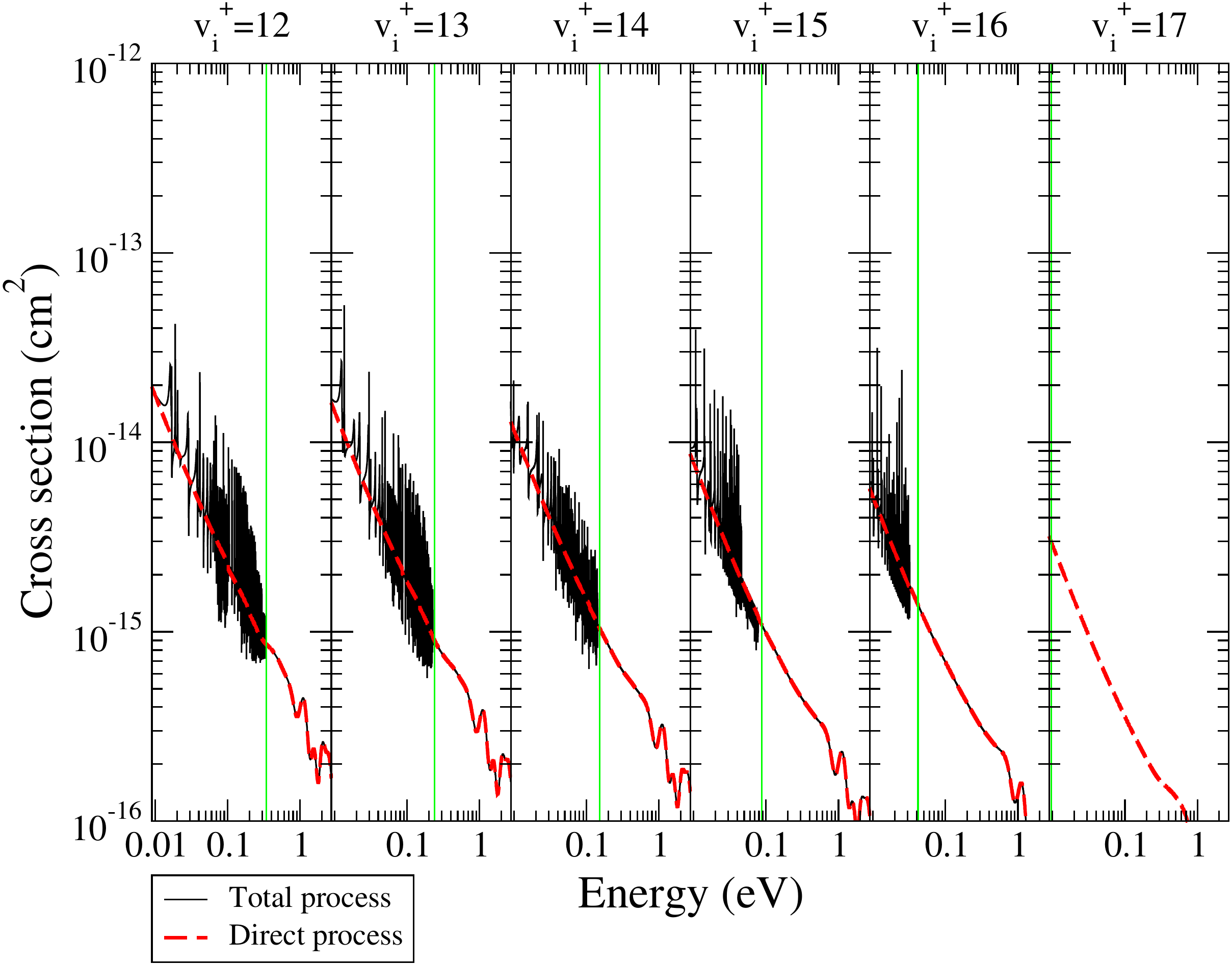} 
\caption{\label{fig:4} Dissociative recombination cross sections of excited ($v_{i}^{+}=12,13,...,17$) {\rm BeH$^{+}$} in its electronic ground state. Direct mechanism: dashed thick line, total (direct and indirect) mechanism: continuos thin line.
}
\end{Dfigures}

\clearpage

 \begin{Dfigures}[h]
\includegraphics[width=0.9\linewidth]{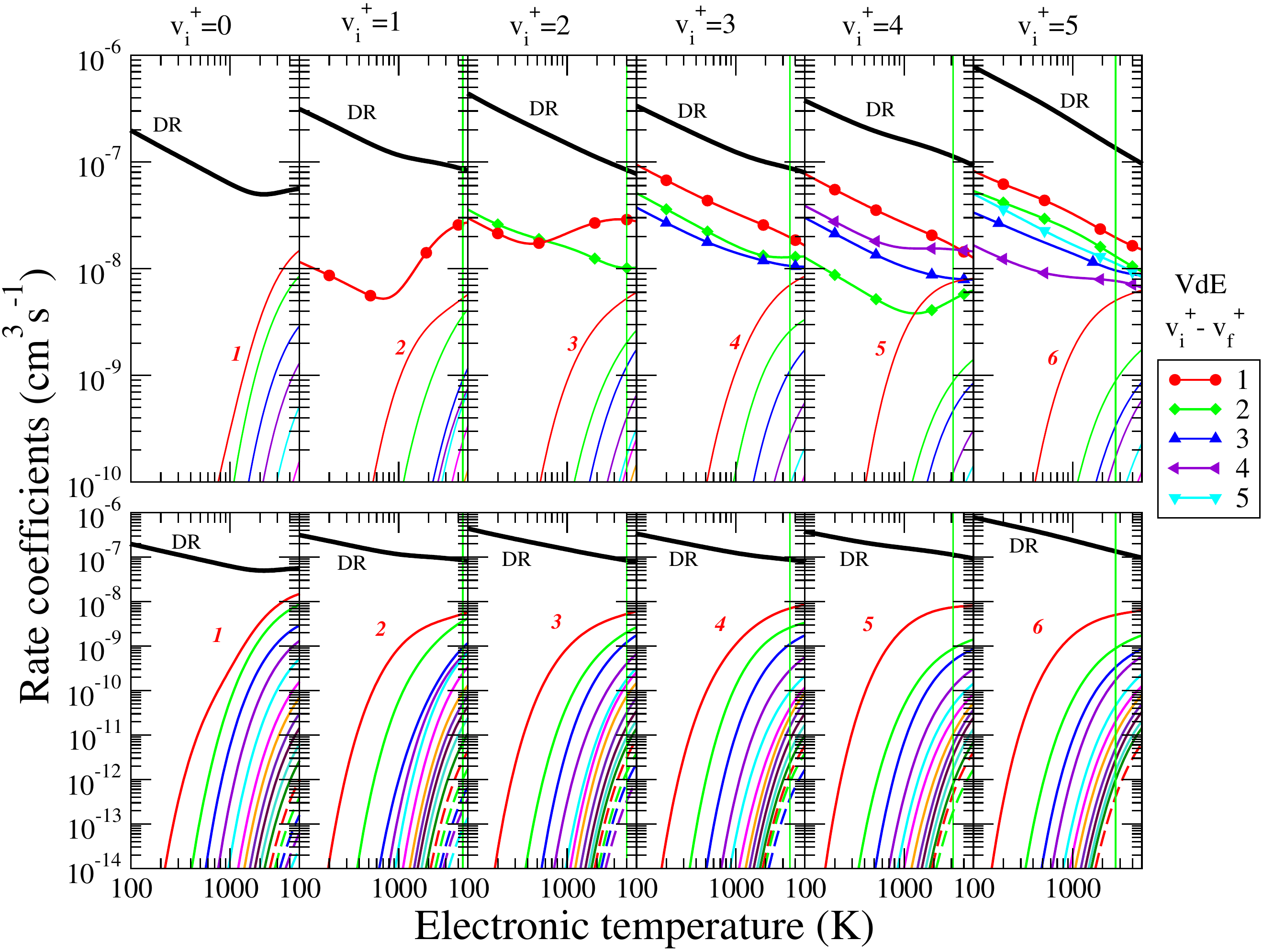} 
\caption{\label{fig:5} Dissociative recombination (DR, thick line), vibrational excitation (VE, thin lines) and vibrational de-excitation (VdE, symbols and thick lines) Maxwell rate coefficients of ground ($v_{i}^{+}=0$) and excited ($v_{i}^{+}=1,...,5$) {\rm BeH$^{+}$} in its electronic ground state (total mechanism). For VE, since the rate coefficients decrease monotonically with the excitation, the lowest final vibrational quantum number of the target is indicated only, and the lower panels extend the range down to \rm{10$^{-14}$ cm$^3$/s}.}
\end{Dfigures}

\clearpage

 \begin{Dfigures}[h]
\includegraphics[width=0.9\linewidth]{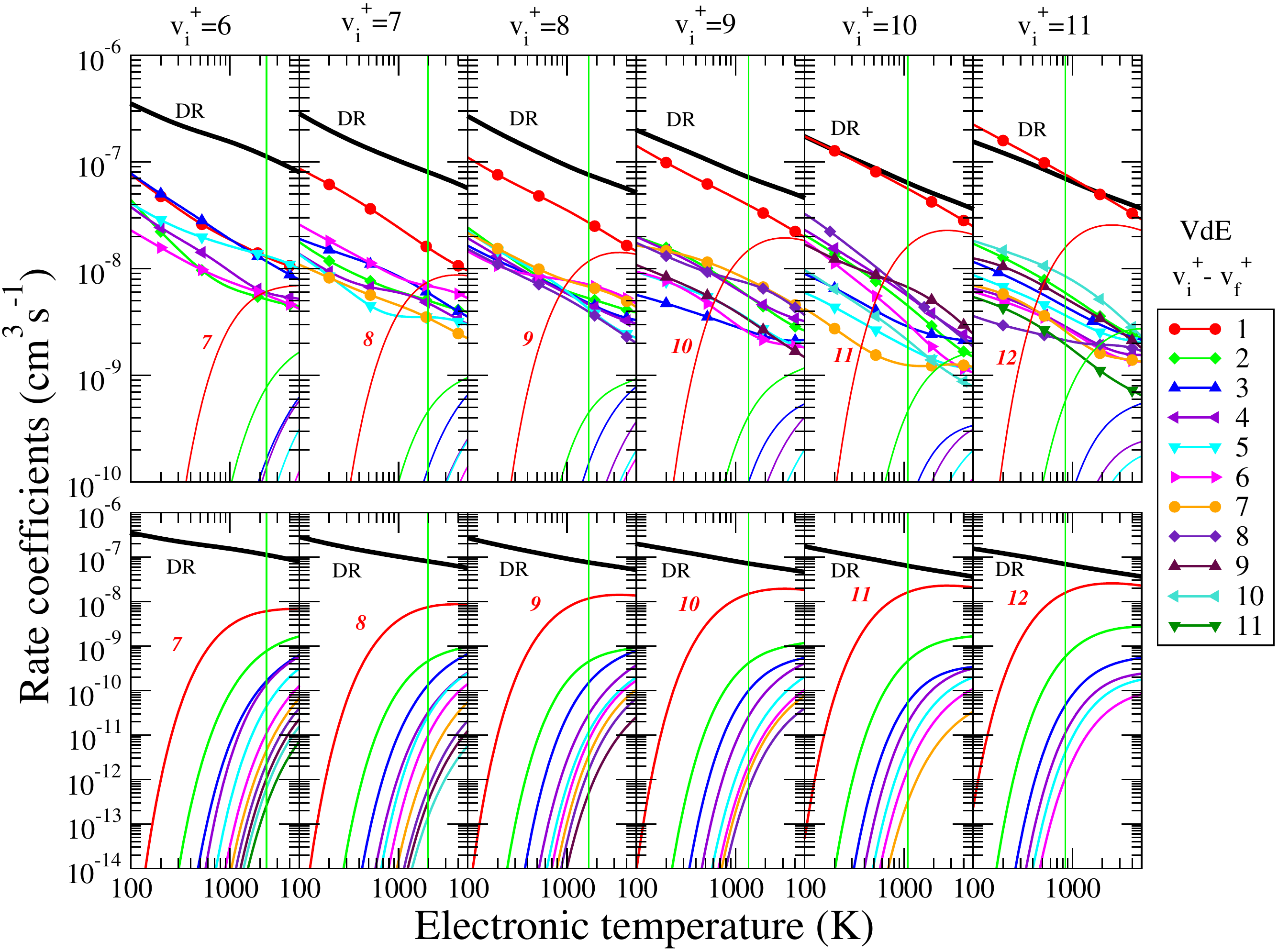}
\caption{\label{fig:6} 
Dissociative recombination (DR, thick line), vibrational excitation (VE, thin lines) and vibrational de-excitation (VdE, symbols and thick lines) Maxwell rate coefficients of excited ($v_{i}^{+}=7,8,...,11$) {\rm BeH$^{+}$} in its electronic ground state (total mechanism). For VE, since the rate coefficients decrease monotonically with the excitation, the lowest final vibrational quantum number of the target is indicated only, and the lower panels extend the range down to \rm{10$^{-14}$ cm$^3$/s}.}
\end{Dfigures}

\clearpage

 \begin{Dfigures}[h]
\includegraphics[width=0.9\linewidth]{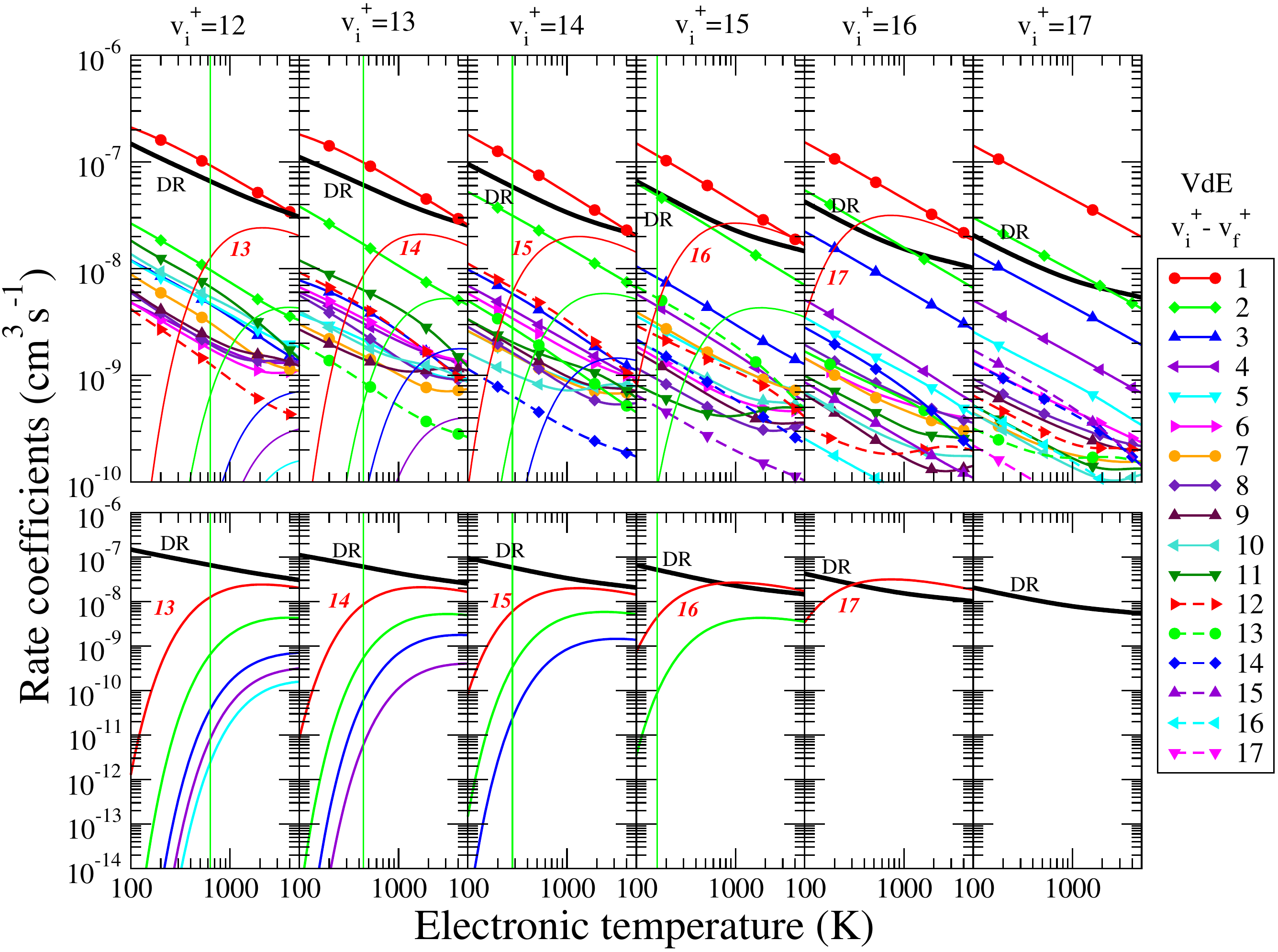}
\caption{\label{fig:7} 
Dissociative recombination (DR, thick line), vibrational excitation (VE, thin lines) and vibrational de-excitation (VdE, symbols and thick lines) Maxwell rate coefficients of excited ($v_{i}^{+}=12,13,...,17$) {\rm BeH$^{+}$} in its electronic ground state (total mechanism). For VE, since the rate coefficients decrease monotonically with the excitation, the lowest final vibrational quantum number of the target is indicated only, and the lower panels extend the range down to \rm{10$^{-14}$ cm$^3$/s}.}
\end{Dfigures}

\clearpage

 \begin{Dfigures}[h]
\includegraphics[width=0.9\linewidth]{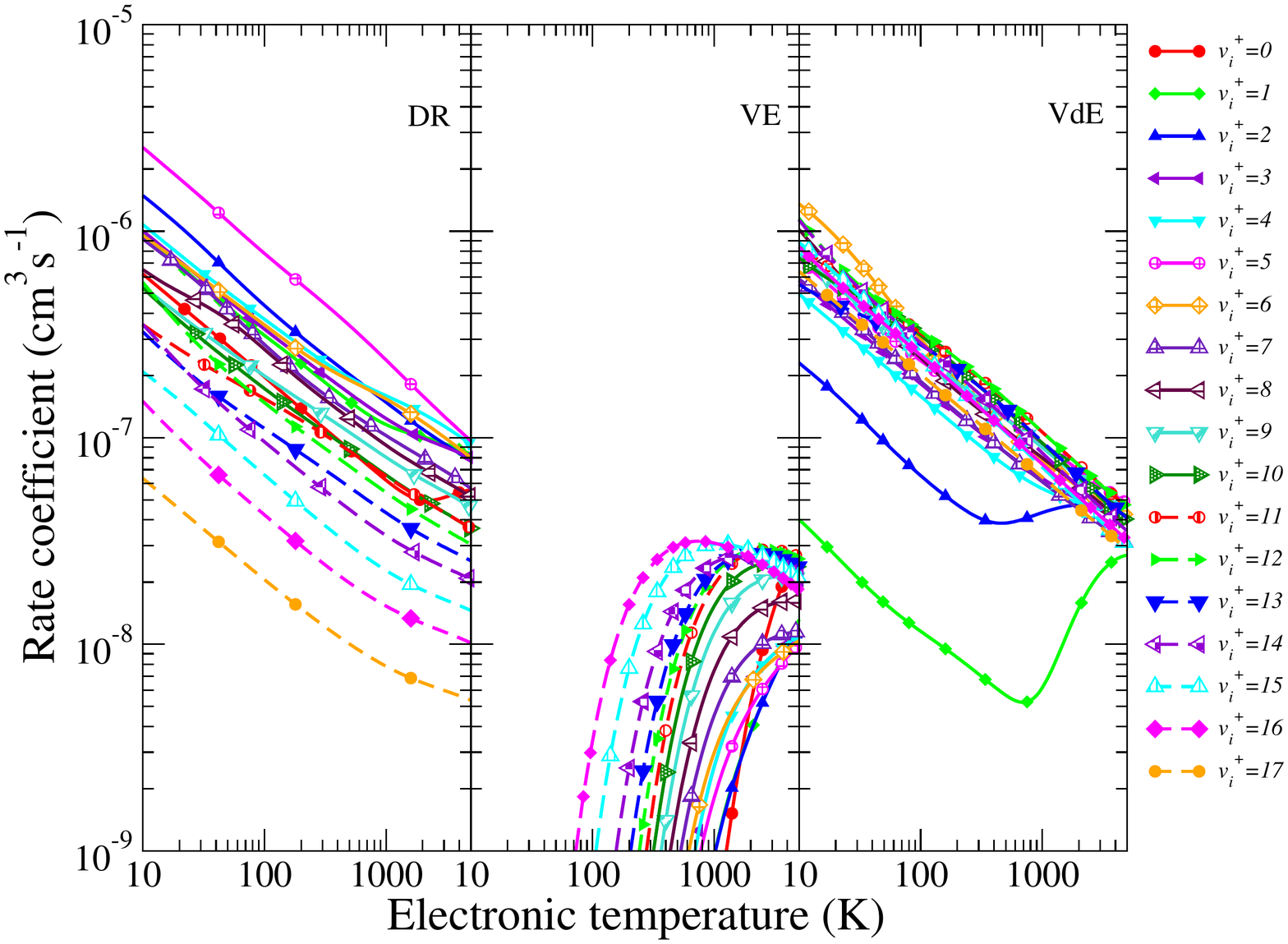}
\caption{\label{fig:8} 
Global (sum over all the possible final states) Maxwell rate coefficients for dissociative recombination (DR), vibrational excitation (VE) and vibrational de-excitation (VdE).} 
\end{Dfigures}

\clearpage

\TableExplanation

\subsection*{Table~\ref{tab:BeH_DR_Interpolation} List of fitting parameters according to eq.~(\ref{eqn:BeH_DR_Interpolation}) with 'P'='DR' and 'L'=$v_i^+$' calculated for dissociative recombination for all vibrational levels of the ground electronic state of BeH$^+$.}
\begin{tabular}{@{}p{2.5in}p{6in}@{}}
$v_i^+$		& Initial vibrational level of BeH$^+$ \\
$A_{v^+_i}$, $\alpha_{v^+_i}$, $B_{v^+_i}(j=1,\dots,7)$	& Fitting parameters \\
\end{tabular}

\subsection*{Tables~\ref{tab:BeH_VE_Interpolation0-1}-\ref{tab:BeH_VE_Interpolation16-17} List of fitting parameters according to eq.~(\ref{eqn:BeH_DR_Interpolation}) with 'P'='VE' or 'VdE' and 'L'=$v_i^+\to v_f^+$' calculated for vibrational transitions (excitation and de-excitation) between the vibrational levels of the ground electronic state of BeH$^+$.}
\begin{tabular}{@{}p{2.5in}p{6in}@{}}
$v_i^+\to v_f^+$		& Vibrational transition of BeH$^+$ \\
 $\boldsymbol{v_i^+}$$\to$$\boldsymbol{ v_f^+}$	& Stands for VdE\\
$A_{v^+_i\to v^+_f}$, $\alpha_{v^+_i\to v^+_f}$, $B_{v^+_i\to v^+_f}(j=1,\dots,7)$	& Fitting parameters \\
\end{tabular}

\clearpage
\datatables

\begin{landscape}
\renewcommand{\thefootnote}{\fnsymbol{footnote}}
\renewcommand{\arraystretch}{0.81}
\scriptsize
\begin{longtable}{lccccccccc}
\caption[table1]{List of the parameters used in Eq.(\ref{eqn:BeH_DR_Interpolation}) for the DR Maxwell rate coefficients of BeH$^+(v_{i}^{+}=0,1,2,3,...,17)$ displayed in Graphs~\ref{fig:5},~\ref{fig:6}, and~\ref{fig:7}.}
\label{tab:BeH_DR_Interpolation}\vspace{0.5cm}\\
\hline \\[-2ex]
   \multicolumn{1}{c}{$v^+_i$} &
   \multicolumn{1}{c}{$A_{v^+_i}$} &
   \multicolumn{1}{c}{$\alpha_{v^+_i}$} &
   \multicolumn{1}{c}{$B_{v^+_i}(1)$} &
   \multicolumn{1}{c}{$B_{v^+_i}(2)$} &
   \multicolumn{1}{c}{$B_{v^+_i}(3)$} &
   \multicolumn{1}{c}{$B_{v^+_i}(4)$}  &
   \multicolumn{1}{c}{$B_{v^+_i}(5)$} &
   \multicolumn{1}{c}{$B_{v^+_i}(6)$} &
   \multicolumn{1}{c}{$B_{v^+_i}(7)$} \\[5pt]
 \hline \\[-2ex]
\endhead
    0   &  0.253513120$\times10^{-9}$ &   0.592469479 & -0.190649410$\times10^{4}$ &  0.123315004$\times10^{7}$  & -0.388342503$\times10^{9}$ & 0.647265585$\times10^{11}$   & -0.566722264$\times10^{13}$ & 0.232068987$\times10^{15}$  & -0.295078806$\times10^{16}$ \\
    1   &  0.314462330$\times10^{-5}$ &  -0.407538250 & 0.958577714$\times10^{3}$  &  -0.135957934$\times10^{7}$ & 0.725436198$\times10^{9}$  & -0.192640935$\times10^{12}$  & 0.271812541$\times10^{14}$  & -0.194765942$\times10^{16}$ & 0.557221814$\times10^{17}$ \\
    2   &  0.114678715$\times10^{-5}$ &  -0.322376187 & -0.265630394$\times10^{3}$ &  0.219601686$\times10^{6}$  & -0.916549130$\times10^{8}$ & 0.207255861$\times10^{11}$   & -0.259613563$\times10^{13}$ & 0.169396874$\times10^{15}$  & -0.449074035$\times10^{16}$ \\
		3   &  0.849641548$\times10^{-6}$ &  -0.275140774 & 0.171673351$\times10^{3}$  &  -0.451956583$\times10^{6}$ & 0.285096525$\times10^{9}$  &  -0.827403486$\times10^{11}$ & 0.123588751$\times10^{14}$  & -0.921680323$\times10^{15}$ & 0.271612346$\times10^{17}$ \\
4   &  .121239628$\times10^{-04}$ &  -.555845997$\times10^{00}$ & .761294556$\times10^{03}$  &  -.714351056$\times10^{06}$ & .302073187$\times10^{09}$  &  -.687621706$\times10^{11}$ & .869454121$\times10^{13}$  &  -.574150868$\times10^{15}$ & .154263413$\times10^{17}$  \\
		5   &  .644139392$\times10^{-05}$ &  -.498481741$\times10^{00}$ & -.238774152$\times10^{03}$ & .250065677$\times10^{06}$   & -.107836783$\times10^{09}$ &  .245547321$\times10^{11}$  & -.309392723$\times10^{13}$ &  .203550809$\times10^{15}$  &  -.545188444$\times10^{16}$ \\
		6   & .792927406$\times10^{-05}$  &  -.532262180$\times10^{00}$ & .333511522$\times10^{03}$  &  -.148501292$\times10^{06}$ & .171708872$\times10^{08}$  &  .305133187$\times10^{10}$  & -.998389272$\times10^{12}$ &  .951736825$\times10^{14}$  &  -.314431818$\times10^{16}$ \\
\end{longtable}
\normalsize
\renewcommand{\thefootnote}{\arabic{footnote}}
\renewcommand{\arraystretch}{1.0}
\end{landscape}

\newpage

\begin{landscape}
\renewcommand{\thefootnote}{\fnsymbol{footnote}}
\renewcommand{\arraystretch}{0.81}
\scriptsize
\setlength\tabcolsep{1.2ex}
\begin{longtable}{lccccccccc}
\caption[table2]{List of the parameters used in Eq.(\ref{eqn:BeH_DR_Interpolation}) for the VE and VdE Maxwell rate coefficients of BeH$^+$  ($v^+_i$~=~0 and 1) represented in the upper panels of Graph~\ref{fig:5}.}\label{tab:BeH_VE_Interpolation0-1}\vspace{0.5cm}\\
\hline \\[-2ex]
   \multicolumn{1}{c}{$v^+_i \to v^+_f$} &
   \multicolumn{1}{c}{$A_{{v^+_i}\to {v^+_f}}$} &
   \multicolumn{1}{c}{$\alpha_{{v^+_i}\to {v^+_f}}$} &
   \multicolumn{1}{c}{$B_{{v^+_i}\to {v^+_f}}(1)$} &
   \multicolumn{1}{c}{$B_{{v^+_i}\to {v^+_f}}(2)$} &
   \multicolumn{1}{c}{$B_{{v^+_i}\to {v^+_f}}(3)$} &
   \multicolumn{1}{c}{$B_{{v^+_i}\to {v^+_f}}(4)$}  &
   \multicolumn{1}{c}{$B_{{v^+_i}\to {v^+_f}}(5)$} &
   \multicolumn{1}{c}{$B_{{v^+_i}\to {v^+_f}}(6)$} &
   \multicolumn{1}{c}{$B_{{v^+_i}\to {v^+_f}}(7)$} \\[5pt]
 \hline \\[-2ex]
\endhead
0 $\to$ 1  & 0.124244699$\times10^{-01}$  & -0.133548647$\times10^{+01}$  & 0.125068733$\times10^{+05}$  & -0.119462867$\times10^{+08}$  & 0.612265338$\times10^{+10}$  & -0.159115695$\times10^{+13}$  & 0.221261867$\times10^{+15}$  & -0.156815975$\times10^{+17}$  & 0.444766919$\times10^{+18}$ \\
 0 $\to$ 2  & 0.129961950$\times10^{-04}$  & -0.661035442$\times10^{+00}$  & 0.875512695$\times10^{+04}$  & -0.266357941$\times10^{+07}$  & 0.108710933$\times10^{+10}$  & -0.242430358$\times10^{+12}$  & 0.306483606$\times10^{+14}$  & -0.205871435$\times10^{+16}$  & 0.569537410$\times10^{+17}$ \\
 0 $\to$ 3  & 0.142379982$\times10^{-04}$  & -0.764830977$\times10^{+00}$  & 0.101101728$\times10^{+05}$  & -0.205253723$\times10^{+07}$  & 0.137208247$\times10^{+10}$  & -0.493346150$\times10^{+12}$  & 0.974091099$\times10^{+14}$  & -0.991580495$\times10^{+16}$  & 0.405560372$\times10^{+18}$ \\
 0 $\to$ 4  & 0.613107973$\times10^{-05}$  & -0.715948096$\times10^{+00}$  & 0.116459658$\times10^{+05}$  & 0.180037625$\times10^{+07}$  & -0.380655693$\times10^{+10}$  & 0.289048070$\times10^{+13}$  & -0.104565345$\times10^{+16}$  & 0.181732642$\times10^{+18}$  & -0.121943739$\times10^{+20}$ \\
0 $\to$  5  & 0.454423919$\times10^{-05}$  & -0.724833034$\times10^{+00}$  & 0.144167691$\times10^{+05}$  & 0.187042671$\times10^{+07}$  & -0.553846921$\times10^{+10}$  & 0.525679261$\times10^{+13}$  & -0.233886685$\times10^{+16}$  & 0.498406730$\times10^{+18}$  & -0.410074688$\times10^{+20}$ \\
 0 $\to$ 6  & 0.779183743$\times10^{-05}$  & -0.847008699$\times10^{+00}$  & 0.181829327$\times10^{+05}$  & -0.181366766$\times10^{+07}$  & -0.132640151$\times10^{+10}$  & 0.318338532$\times10^{+13}$  & -0.206370313$\times10^{+16}$  & 0.581977475$\times10^{+18}$  & -0.613361897$\times10^{+20}$ \\
\end{longtable}
\normalsize
\renewcommand{\thefootnote}{\arabic{footnote}}
\renewcommand{\arraystretch}{1.0}
\end{landscape}

\newpage
\begin{landscape}
\renewcommand{\thefootnote}{\fnsymbol{footnote}}
\renewcommand{\arraystretch}{0.81}
\scriptsize
\setlength\tabcolsep{1.2ex}
\begin{longtable}{lccccccccc}
\caption[table3]{List of the parameters used in Eq.(\ref{eqn:BeH_DR_Interpolation}) for the VE and VdE Maxwell rate coefficients of BeH$^+$  ($v^+_i$~=~2 and 3) represented in the upper panels of Graph~\ref{fig:5}.} \label{tab:BeH_VE_Interpolation2-3}\vspace{0.5cm}\\
\hline \\[-2ex]
   \multicolumn{1}{c}{$v^+_i \to v^+_f$} &
   \multicolumn{1}{c}{$A_{{v^+_i}\to {v^+_f}}$} &
   \multicolumn{1}{c}{$\alpha_{{v^+_i}\to {v^+_f}}$} &
   \multicolumn{1}{c}{$B_{{v^+_i}\to {v^+_f}}(1)$} &
   \multicolumn{1}{c}{$B_{{v^+_i}\to {v^+_f}}(2)$} &
   \multicolumn{1}{c}{$B_{{v^+_i}\to {v^+_f}}(3)$} &
   \multicolumn{1}{c}{$B_{{v^+_i}\to {v^+_f}}(4)$}  &
   \multicolumn{1}{c}{$B_{{v^+_i}\to {v^+_f}}(5)$} &
   \multicolumn{1}{c}{$B_{{v^+_i}\to {v^+_f}}(6)$} &
   \multicolumn{1}{c}{$B_{{v^+_i}\to {v^+_f}}(7)$} \\[5pt]
 \hline \\[-2ex]
\endhead
{\bf 2 $\to$ 0}  & 0.126308415$\times10^{-09}$  & 0.456305622$\times10^{+00}$  & -0.282698010$\times10^{+04}$  & 0.324128858$\times10^{+07}$  & -0.166236562$\times10^{+10}$  & 0.438820087$\times10^{+12}$  & -0.620881147$\times10^{+14}$  & 0.447096361$\times10^{+16}$  & -0.128575131$\times10^{+18}$ \\
{\bf 2 $\to$ 1}  & 0.108956019$\times10^{-04}$  & -0.641702757$\times10^{+00}$  & 0.276723409$\times10^{+04}$  & -0.249602163$\times10^{+07}$  & 0.977115209$\times10^{+09}$  & -0.206865990$\times10^{+12}$  & 0.246613751$\times10^{+14}$  & -0.155686663$\times10^{+16}$  & 0.404532125$\times10^{+17}$ \\
2 $\to$ 3  & 0.296926175$\times10^{-08}$  & 0.124816242$\times10^{+00}$  & 0.173836549$\times10^{+04}$  & 0.908486506$\times10^{+06}$  & -0.424294717$\times10^{+09}$  & 0.108080533$\times10^{+12}$  & -0.150733302$\times10^{+14}$  & 0.107908466$\times10^{+16}$  & -0.309653342$\times10^{+17}$ \\
2 $\to$ 4  & 0.233152145$\times10^{-08}$  & 0.122367759$\times10^{+00}$  & 0.452862967$\times10^{+04}$  & 0.653277887$\times10^{+06}$  & -0.273049775$\times10^{+09}$  & 0.670734633$\times10^{+11}$  & -0.944014877$\times10^{+13}$  & 0.700586130$\times10^{+15}$  & -0.211863471$\times10^{+17}$ \\
2 $\to$ 5  & 0.179746490$\times10^{-09}$  & 0.385415212$\times10^{+00}$  & 0.455113975$\times10^{+04}$  & 0.557366075$\times10^{+07}$  & -0.437760675$\times10^{+10}$  & 0.181187931$\times10^{+13}$  & -0.404354650$\times10^{+15}$  & 0.458962441$\times10^{+17}$  & -0.207392120$\times10^{+19}$ \\
2 $\to$ 6  & 0.662229481$\times10^{-13}$  & 0.115712028$\times10^{+01}$  & 0.161972309$\times10^{+04}$  & 0.182461133$\times10^{+08}$  & -0.191412396$\times10^{+11}$  & 0.106134357$\times10^{+14}$  & -0.314224909$\times10^{+16}$  & 0.468199997$\times10^{+18}$  & -0.275398722$\times10^{+20}$ \\
\end{longtable}
\normalsize
\renewcommand{\thefootnote}{\arabic{footnote}}
\renewcommand{\arraystretch}{1.0}
\end{landscape}

\newpage

\begin{landscape}
\renewcommand{\thefootnote}{\fnsymbol{footnote}}
\renewcommand{\arraystretch}{0.81}
\scriptsize
\setlength\tabcolsep{1.2ex}
\begin{longtable}{lccccccccc}
\caption[table4]{List of the parameters used in Eq.(\ref{eqn:BeH_DR_Interpolation}) for the VE and VdE Maxwell rate coefficients of BeH$^+$  ($v^+_i$~=~4 and 5) represented in the upper panels of Graph~\ref{fig:5}.}\label{tab:BeH_VE_Interpolation4-5}\vspace{0.5cm}\\
\hline \\[-2ex]
   \multicolumn{1}{c}{$v^+_i \to v^+_f$} &
   \multicolumn{1}{c}{$A_{{v^+_i}\to {v^+_f}}$} &
   \multicolumn{1}{c}{$\alpha_{{v^+_i}\to {v^+_f}}$} &
   \multicolumn{1}{c}{$B_{{v^+_i}\to {v^+_f}}(1)$} &
   \multicolumn{1}{c}{$B_{{v^+_i}\to {v^+_f}}(2)$} &
   \multicolumn{1}{c}{$B_{{v^+_i}\to {v^+_f}}(3)$} &
   \multicolumn{1}{c}{$B_{{v^+_i}\to {v^+_f}}(4)$}  &
   \multicolumn{1}{c}{$B_{{v^+_i}\to {v^+_f}}(5)$} &
   \multicolumn{1}{c}{$B_{{v^+_i}\to {v^+_f}}(6)$} &
   \multicolumn{1}{c}{$B_{{v^+_i}\to {v^+_f}}(7)$} \\[5pt]
 \hline \\[-2ex]
\endhead
{\bf 4~$\to$~ 0}  &   .358062377$\times10^{-07}$ & -.111707819$\times10^{00}$ & -.523486685$\times10^{03}$ &  .372738921$\times10^{07}$ & -.678756195$\times10^{10}$ &  .539351308$\times10^{13}$ & -.218437908$\times10^{16}$ &  .443605968$\times10^{18}$ & -.358959655$\times10^{20}$ \\
{\bf 4~$\to$~ 1}  &   .604013983$\times10^{-09}$ &  .265464781$\times10^{00}$ & -.175674106$\times10^{04}$ &  .244331137$\times10^{07}$ & -.200984104$\times10^{10}$ &  .931164967$\times10^{12}$ & -.231463009$\times10^{15}$ &  .271464752$\times10^{17}$ & -.992843627$\times10^{18}$ \\
{\bf 4~$\to$~ 2}  &   .623253104$\times10^{-03}$ & -.112985446$\times10^{01}$ &  .121940642$\times10^{05}$ & -.321116064$\times10^{08}$ &  .376208253$\times10^{11}$ & -.237896655$\times10^{14}$ &  .841392067$\times10^{16}$ & -.156579408$\times10^{19}$ &  .119370657$\times10^{21}$ \\
{\bf 4~$\to$~ 3}  &   .160723340$\times10^{-03}$ & -.104889329$\times10^{01}$ &  .314477056$\times10^{04}$ & -.619056009$\times10^{07}$ &  .640948266$\times10^{10}$ & -.381013560$\times10^{13}$ &  .130237723$\times10^{16}$ & -.237528849$\times10^{18}$ &  .178829597$\times10^{20}$ \\
 4~$\to$~ 5  &   .110993265$\times10^{-07}$ & -.359602680$\times10^{-01}$ & -.506031232$\times10^{03}$ &  .660311514$\times10^{07}$ & -.716707150$\times10^{10}$ &  .440032207$\times10^{13}$ & -.154246121$\times10^{16}$ &  .286977165$\times10^{18}$ & -.219488338$\times10^{20}$ \\
 4~$\to$~ 6  &   .142770506$\times10^{-09}$ &  .305813247$\times10^{00}$ &  .645344607$\times10^{03}$ &  .111449971$\times10^{08}$ & -.144722005$\times10^{11}$ &  .981325730$\times10^{13}$ & -.361377623$\times10^{16}$ &  .687380153$\times10^{18}$ & -.529733459$\times10^{20}$ \\
 4~$\to$~ 7  &   .285584691$\times10^{-07}$ & -.283108254$\times10^{00}$ &  .509946612$\times10^{04}$ &  .492954781$\times10^{07}$ & -.552991649$\times10^{10}$ &  .343706995$\times10^{13}$ & -.121035857$\times10^{16}$ &  .225460627$\times10^{18}$ & -.172357375$\times10^{20}$ \\
\end{longtable}
\normalsize
\renewcommand{\thefootnote}{\arabic{footnote}}
\renewcommand{\arraystretch}{1.0}
\end{landscape}

\newpage

\begin{landscape}
\renewcommand{\thefootnote}{\fnsymbol{footnote}}
\renewcommand{\arraystretch}{0.81}
\scriptsize
\setlength\tabcolsep{1.2ex}
\begin{longtable}{lccccccccc}
\caption[table5]{List of the parameters used in Eq.(\ref{eqn:BeH_DR_Interpolation}) for the VE and VdE Maxwell rate coefficients of BeH$^+$  ($v^+_i$~=~6 and 7) represented in the upper panels of Graph~\ref{fig:6}.}\label{tab:BeH_VE_Interpolation6-7}\vspace{0.5cm}\\
\hline \\[-2ex]
   \multicolumn{1}{c}{$v^+_i \to v^+_f$} &
   \multicolumn{1}{c}{$A_{{v^+_i}\to {v^+_f}}$} &
   \multicolumn{1}{c}{$\alpha_{{v^+_i}\to {v^+_f}}$} &
   \multicolumn{1}{c}{$B_{{v^+_i}\to {v^+_f}}(1)$} &
   \multicolumn{1}{c}{$B_{{v^+_i}\to {v^+_f}}(2)$} &
   \multicolumn{1}{c}{$B_{{v^+_i}\to {v^+_f}}(3)$} &
   \multicolumn{1}{c}{$B_{{v^+_i}\to {v^+_f}}(4)$}  &
   \multicolumn{1}{c}{$B_{{v^+_i}\to {v^+_f}}(5)$} &
   \multicolumn{1}{c}{$B_{{v^+_i}\to {v^+_f}}(6)$} &
   \multicolumn{1}{c}{$B_{{v^+_i}\to {v^+_f}}(7)$} \\[5pt]
 \hline \\[-2ex]
\endhead
{\bf 6~$\to$~ 0}  &  .101619967$\times10^{-05}$ & -.617687637$\times10^{00}$ &  .134770508$\times10^{04}$ & -.245027759$\times10^{07}$ &  .232350563$\times10^{10}$ & -.134466407$\times10^{13}$ &  .468395019$\times10^{15}$ & -.890228309$\times10^{17}$ &  .703466728$\times10^{19}$ \\
{\bf 6~$\to$~ 1}  &  .265753802$\times10^{-02}$ & -.134820531$\times10^{01}$ &  .632328744$\times10^{04}$ & -.140168179$\times10^{08}$ &  .157316976$\times10^{11}$ & -.985590430$\times10^{13}$ &  .348499630$\times10^{16}$ & -.649831258$\times10^{18}$ &  .496519521$\times10^{20}$ \\
{\bf 6~$\to$~ 2}  &  .880428656$\times10^{-06}$ & -.527069495$\times10^{00}$ &  .468191103$\times10^{04}$ & -.178472039$\times10^{08}$ &  .256601637$\times10^{11}$ & -.185117478$\times10^{14}$ &  .717228088$\times10^{16}$ & -.142607210$\times10^{19}$ &  .114268616$\times10^{21}$ \\
{\bf 6~$\to$~ 3}  &  .883710078$\times10^{-07}$ & -.310062285$\times10^{00}$ & -.139559798$\times10^{04}$ &  .335336275$\times10^{07}$ & -.431735790$\times10^{10}$ &  .303333006$\times10^{13}$ & -.116883363$\times10^{16}$ &  .232275162$\times10^{18}$ & -.186163656$\times10^{20}$ \\
{\bf 6~$\to$~ 4}  &  .809018782$\times10^{-11}$ &  .633893753$\times10^{00}$ & -.583317334$\times10^{04}$ &  .147984046$\times10^{08}$ & -.188219571$\times10^{11}$ &  .126982896$\times10^{14}$ & -.468766659$\times10^{16}$ &  .896837888$\times10^{18}$ & -.696135587$\times10^{20}$ \\
{\bf 6~$\to$~ 5}  &  .241670141$\times10^{-09}$ &  .358475784$\times10^{00}$ & -.414106724$\times10^{04}$ &  .848923943$\times10^{07}$ & -.899468181$\times10^{10}$ &  .531262744$\times10^{13}$ & -.177839346$\times10^{16}$ &  .316008073$\times10^{18}$ & -.231644863$\times10^{20}$ \\
 6~$\to$~ 7  &  .879968218$\times10^{-08}$ & -.274854872$\times10^{-01}$ & -.398362807$\times10^{03}$ &  .455605247$\times10^{07}$ & -.417876209$\times10^{10}$ &  .223431647$\times10^{13}$ & -.698025120$\times10^{15}$ &  .118046686$\times10^{18}$ & -.834216079$\times10^{19}$ \\
\end{longtable}
\normalsize
\renewcommand{\thefootnote}{\arabic{footnote}}
\renewcommand{\arraystretch}{1.0}
\end{landscape}

\newpage

\begin{landscape}
\renewcommand{\thefootnote}{\fnsymbol{footnote}}
\renewcommand{\arraystretch}{0.81}
\scriptsize
\setlength\tabcolsep{1.2ex}
\begin{longtable}{lccccccccc}
\caption[table6]{List of the parameters used in Eq.(\ref{eqn:BeH_DR_Interpolation}) for the VE and VdE Maxwell rate coefficients of BeH$^+$  ($v^+_i$~=~8 and 9) represented in the upper panels of Graph~\ref{fig:6}.}\label{tab:BeH_VE_Interpolation8-9}\vspace{0.5cm}\\
\hline \\[-2ex]
   \multicolumn{1}{c}{$v^+_i \to v^+_f$} &
   \multicolumn{1}{c}{$A_{{v^+_i}\to {v^+_f}}$} &
   \multicolumn{1}{c}{$\alpha_{{v^+_i}\to {v^+_f}}$} &
   \multicolumn{1}{c}{$B_{{v^+_i}\to {v^+_f}}(1)$} &
   \multicolumn{1}{c}{$B_{{v^+_i}\to {v^+_f}}(2)$} &
   \multicolumn{1}{c}{$B_{{v^+_i}\to {v^+_f}}(3)$} &
   \multicolumn{1}{c}{$B_{{v^+_i}\to {v^+_f}}(4)$}  &
   \multicolumn{1}{c}{$B_{{v^+_i}\to {v^+_f}}(5)$} &
   \multicolumn{1}{c}{$B_{{v^+_i}\to {v^+_f}}(6)$} &
   \multicolumn{1}{c}{$B_{{v^+_i}\to {v^+_f}}(7)$} \\[5pt]
 \hline \\[-2ex]
\endhead
{\bf 8~$\to$~ 0}  &  .184693578$\times10^{-06}$ &  -.541917857$\times10^{ 00}$ &  -.594213189$\times10^{ 03}$ &   .148190306$\times10^{ 07}$ &  -.132630145$\times10^{ 10}$ &   .582145197$\times10^{ 12}$ &  -.127196542$\times10^{ 15}$ &   .113530130$\times10^{ 17}$ &  -.800511948$\times10^{ 17}$ \\
{\bf 8~$\to$~ 1}  &  .323058686$\times10^{-03}$ &  -.123015455$\times10^{ 01}$ &   .430770114$\times10^{ 04}$ &  -.709124424$\times10^{ 07}$ &   .572452855$\times10^{ 10}$ &  -.258979176$\times10^{ 13}$ &   .667954033$\times10^{ 15}$ &  -.913986871$\times10^{ 17}$ &   .511609551$\times10^{ 19}$ \\
{\bf 8~$\to$~ 2}  &  .134008508$\times10^{-03}$ &  -.112344155$\times10^{ 01}$ &   .415557459$\times10^{ 04}$ &  -.816462104$\times10^{ 07}$ &   .900723764$\times10^{ 10}$ &  -.574711839$\times10^{ 13}$ &   .208695008$\times10^{ 16}$ &  -.399497593$\times10^{ 18}$ &   .312484365$\times10^{ 20}$ \\
{\bf 8~$\to$~ 3}  &  .605666613$\times10^{-10}$ &   .334311293$\times10^{ 00}$ &  -.441645243$\times10^{ 04}$ &   .639486991$\times10^{ 07}$ &  -.400570022$\times10^{ 10}$ &   .993028253$\times10^{ 12}$ &   .610314247$\times10^{ 14}$ &  -.718490612$\times10^{ 17}$ &   .920389754$\times10^{ 19}$ \\
{\bf 8~$\to$~ 4} &  .156474286$\times10^{-09}$ &   .275016049$\times10^{ 00}$ &  -.383540503$\times10^{ 04}$ &   .794359625$\times10^{ 07}$ &  -.847158072$\times10^{ 10}$ &   .505189062$\times10^{ 13}$ &  -.170596928$\times10^{ 16}$ &   .305050575$\times10^{ 18}$ &  -.224454503$\times10^{ 20}$ \\
{\bf 8~$\to$~ 5}  &  .165340301$\times10^{-09}$ &   .287386106$\times10^{ 00}$ &  -.321930128$\times10^{ 04}$ &   .556874013$\times10^{ 07}$ &  -.516219500$\times10^{ 10}$ &   .281259111$\times10^{ 13}$ &  -.898137202$\times10^{ 15}$ &   .155100090$\times10^{ 18}$ &  -.111617933$\times10^{ 20}$ \\
\end{longtable}
\normalsize
\renewcommand{\thefootnote}{\arabic{footnote}}
\renewcommand{\arraystretch}{1.0}
\end{landscape}

\newpage

\begin{landscape}
\renewcommand{\thefootnote}{\fnsymbol{footnote}}
\renewcommand{\arraystretch}{0.81}
\scriptsize
\setlength\tabcolsep{1.2ex}
\begin{longtable}{lccccccccc}
\caption[table7]{List of the parameters used in Eq.~(10) for the VE and VdE Maxwell rate coefficients of BeH$^+$  ($v^+_i$~=~10 and 11) represented in the upper panels of Graph.~5..}\label{tab:BeH_VE_Interpolation10-11}\vspace{0.5cm}\\
\hline \\[-2ex]
   \multicolumn{1}{c}{$v^+_i \to v^+_f$} &
   \multicolumn{1}{c}{$A_{{v^+_i}\to {v^+_f}}$} &
   \multicolumn{1}{c}{$\alpha_{{v^+_i}\to {v^+_f}}$} &
   \multicolumn{1}{c}{$B_{{v^+_i}\to {v^+_f}}(1)$} &
   \multicolumn{1}{c}{$B_{{v^+_i}\to {v^+_f}}(2)$} &
   \multicolumn{1}{c}{$B_{{v^+_i}\to {v^+_f}}(3)$} &
   \multicolumn{1}{c}{$B_{{v^+_i}\to {v^+_f}}(4)$}  &
   \multicolumn{1}{c}{$B_{{v^+_i}\to {v^+_f}}(5)$} &
   \multicolumn{1}{c}{$B_{{v^+_i}\to {v^+_f}}(6)$} &
   \multicolumn{1}{c}{$B_{{v^+_i}\to {v^+_f}}(7)$} \\[5pt]
 \hline \\[-2ex]
\endhead
{\bf 10~$\to$~ 0 }  &   .239260000$\times10^{-08}$ & -.166440996$\times10^{ 00}$ & -.158217399$\times10^{ 04}$ &  .306325853$\times10^{ 06}$ &  .242627010$\times10^{ 10}$ & -.279953761$\times10^{ 13}$ &  .133114147$\times10^{ 16}$ & -.297670872$\times10^{ 18}$ & .257568692$\times10^{ 20}$ \\
{\bf 10~$\to$~ 1}   &   .200920809$\times10^{-04}$ & -.104032246$\times10^{ 01}$ &  .672529357$\times10^{ 03}$ &  .122771792$\times10^{ 07}$ & -.260845095$\times10^{ 10}$ &  .199582919$\times10^{ 13}$ & -.776633465$\times10^{ 15}$ &  .153200280$\times10^{ 18}$ & -.121527192$\times10^{ 20}$ \\
{\bf 10~$\to$~ 2}   &   .253113700$\times10^{-04}$ & -.107473411$\times10^{ 01}$ &  .235961091$\times10^{ 04}$ & -.604270333$\times10^{ 07}$ &  .729911578$\times10^{ 10}$ & -.473924256$\times10^{ 13}$ &  .170874712$\times10^{ 16}$ & -.322545650$\times10^{ 18}$ &  .248612252$\times10^{ 20}$ \\
{\bf 10~$\to$~ 3}   &  .241112388$\times10^{-03}$ & -.127906340$\times10^{ 01}$ &  .818994846$\times10^{ 04}$ & -.184785931$\times10^{ 08}$ &  .201517061$\times10^{ 11}$ & -.122904846$\times10^{ 14}$ &  .425629668$\times10^{ 16}$ & -.780786058$\times10^{ 18}$ &  .588689596$\times10^{ 20}$ \\
{\bf 10~$\to$~ 4}   &   .594023423$\times10^{-11}$ &  .501656971$\times10^{ 00}$ & -.521947559$\times10^{ 04}$ &  .694513952$\times10^{ 07}$ & -.384539190$\times10^{ 10}$ &  .568077951$\times10^{ 12}$ &  .291714829$\times10^{ 15}$ & -.125708146$\times10^{ 18}$ & .139353132$\times10^{ 20}$ \\
{\bf 10~$\to$~ 5}   &   .107506833$\times10^{-09}$ &  .240313266$\times10^{ 00}$ & -.173221041$\times10^{ 04}$ &  .694820600$\times10^{ 06}$ &  .150356129$\times10^{ 10}$ & -.197192276$\times10^{ 13}$ & .974510463$\times10^{ 15}$ & -.222430648$\times10^{ 18}$ &  .195026894$\times10^{ 20}$ \\
\end{longtable}
\normalsize
\renewcommand{\thefootnote}{\arabic{footnote}}
\renewcommand{\arraystretch}{1.0}
\end{landscape}

\newpage

\begin{landscape}
\renewcommand{\thefootnote}{\fnsymbol{footnote}}
\renewcommand{\arraystretch}{0.81}
\scriptsize
\setlength\tabcolsep{1.2ex}
\begin{longtable}{lccccccccc}
\caption[table8]{List of the parameters used in Eq.(\ref{eqn:BeH_DR_Interpolation}) for the VE and VdE Maxwell rate coefficients of BeH$^+$  ($v^+_i$~=~12 and 13) represented in the upper panels of Graph~\ref{fig:7}.}\label{tab:BeH_VE_Interpolation12-13}\vspace{0.5cm}\\
\hline \\[-2ex]
   \multicolumn{1}{c}{$v^+_i \to v^+_f$} &
   \multicolumn{1}{c}{$A_{{v^+_i}\to {v^+_f}}$} &
   \multicolumn{1}{c}{$\alpha_{{v^+_i}\to {v^+_f}}$} &
   \multicolumn{1}{c}{$B_{{v^+_i}\to {v^+_f}}(1)$} &
   \multicolumn{1}{c}{$B_{{v^+_i}\to {v^+_f}}(2)$} &
   \multicolumn{1}{c}{$B_{{v^+_i}\to {v^+_f}}(3)$} &
   \multicolumn{1}{c}{$B_{{v^+_i}\to {v^+_f}}(4)$}  &
   \multicolumn{1}{c}{$B_{{v^+_i}\to {v^+_f}}(5)$} &
   \multicolumn{1}{c}{$B_{{v^+_i}\to {v^+_f}}(6)$} &
   \multicolumn{1}{c}{$B_{{v^+_i}\to {v^+_f}}(7)$} \\[5pt]
 \hline \\[-2ex]
\endhead
{\bf 12~$\to$~ 0}   &   .187042454$\times10^{-07}$ & -.434244052$\times10^{ 00}$ &  .129070140$\times10^{ 04}$ & -.645595864$\times10^{ 07}$ &  .988158345$\times10^{ 10}$ & -.735700671$\times10^{ 13}$ &  .291366015$\times10^{ 16}$ & -.589297290$\times10^{ 18}$ &  .478758481$\times10^{ 20}$ \\
{\bf 12~$\to$~ 1}   &   .218546493$\times10^{-05}$ & -.862575293$\times10^{ 00}$ & -.156338304$\times10^{ 03}$ &  .119753753$\times10^{ 07}$ & -.160555811$\times10^{ 10}$ &  .109090375$\times10^{ 13}$ & -.408485741$\times10^{ 15}$ &  .797233097$\times10^{ 17}$ & -.632365126$\times10^{ 19}$ \\
{\bf 12~$\to$~ 2}   &   .156439995$\times10^{-04}$ & -.106960560$\times10^{ 01}$ &  .830822927$\times10^{ 03}$ &  .708137951$\times10^{ 06}$ & -.219644288$\times10^{ 10}$ &  .186975782$\times10^{ 13}$ & -.771866768$\times10^{ 15}$ &  .158237466$\times10^{ 18}$ & -.128970072$\times10^{ 20}$ \\
{\bf 12~$\to$~ 3}   &   .204729003$\times10^{-03}$ & -.128762785$\times10^{ 01}$ &  .652952108$\times10^{ 04}$ & -.139508435$\times10^{ 08}$ &  .142323899$\times10^{ 11}$ & -.808584851$\times10^{ 13}$ &  .262454603$\times10^{ 16}$ & -.455567245$\times10^{ 18}$ &  .328070303$\times10^{ 20}$ \\
{\bf 12~$\to$~ 4}   &   .356318115$\times10^{-02}$ & -.154173331$\times10^{ 01}$ &  .107318768$\times10^{ 05}$ & -.265858922$\times10^{ 08}$ &  .305484919$\times10^{ 11}$ & -.191159723$\times10^{ 14}$ &  .670564709$\times10^{ 16}$ & -.123864929$\times10^{ 19}$ &  .937772879$\times10^{ 20}$ \\
{\bf 12~$\to$~ 5}   &   .464621774$\times10^{-09}$ &  .859524853$\times10^{-01}$ & -.208259772$\times10^{ 03}$ & -.597458539$\times10^{ 07}$ &  .119104614$\times10^{ 11}$ & -.986002756$\times10^{ 13}$ &  .413511505$\times10^{ 16}$ & -.866485526$\times10^{ 18}$ &  .720884936$\times10^{ 20}$ \\
\end{longtable}
\normalsize
\renewcommand{\thefootnote}{\arabic{footnote}}
\renewcommand{\arraystretch}{1.0}
\end{landscape}

\newpage

\begin{landscape}
\renewcommand{\thefootnote}{\fnsymbol{footnote}}
\renewcommand{\arraystretch}{0.81}
\scriptsize
\setlength\tabcolsep{1.2ex}
\begin{longtable}{lccccccccc}
\caption[table9]{List of the parameters used in Eq.(\ref{eqn:BeH_DR_Interpolation}) for the VE and VdE Maxwell rate coefficients of BeH$^+$  ($v^+_i$~=~14 and 15) represented in the upper panels of Graph~\ref{fig:7}.}\label{tab:BeH_VE_Interpolation14-15}\vspace{0.5cm}\\
\hline \\[-2ex]
   \multicolumn{1}{c}{$v^+_i \to v^+_f$} &
   \multicolumn{1}{c}{$A_{{v^+_i}\to {v^+_f}}$} &
   \multicolumn{1}{c}{$\alpha_{{v^+_i}\to {v^+_f}}$} &
   \multicolumn{1}{c}{$B_{{v^+_i}\to {v^+_f}}(1)$} &
   \multicolumn{1}{c}{$B_{{v^+_i}\to {v^+_f}}(2)$} &
   \multicolumn{1}{c}{$B_{{v^+_i}\to {v^+_f}}(3)$} &
   \multicolumn{1}{c}{$B_{{v^+_i}\to {v^+_f}}(4)$}  &
   \multicolumn{1}{c}{$B_{{v^+_i}\to {v^+_f}}(5)$} &
   \multicolumn{1}{c}{$B_{{v^+_i}\to {v^+_f}}(6)$} &
   \multicolumn{1}{c}{$B_{{v^+_i}\to {v^+_f}}(7)$} \\[5pt]
 \hline \\[-2ex]
\endhead
{\bf 14~$\to$~ 0}   &   .746501579$\times10^{-07}$ & -.663810453$\times10^{ 00}$ &  .272685985$\times10^{ 04}$ & -.791362284$\times10^{ 07}$ &  .101716673$\times10^{ 11}$ & -.698198911$\times10^{ 13}$ &  .264496539$\times10^{ 16}$ & -.520698183$\times10^{ 18}$ &  .415595050$\times10^{ 20}$ \\
{\bf 14~$\to$~ 1}   &   .648276362$\times10^{-07}$ & -.588342904$\times10^{ 00}$ & -.238045874$\times10^{ 03}$ &  .367576608$\times10^{ 06}$ & -.316838382$\times10^{ 09}$ &  .172543203$\times10^{ 12}$ & -.560440472$\times10^{ 14}$ &  .978769015$\times10^{ 16}$ & -.704828271$\times10^{ 18}$ \\
{\bf 14~$\to$~ 2}   &   .861241840$\times10^{-05}$ & -.107872527$\times10^{ 01}$ &  .787319623$\times10^{ 02}$ &  .200368835$\times10^{ 07}$ & -.334888956$\times10^{ 10}$ &  .250143691$\times10^{ 13}$ & -.981945492$\times10^{ 15}$ &  .196631318$\times10^{ 18}$ & -.158329023$\times10^{ 20}$ \\
{\bf 14~$\to$~ 3}   &   .890745457$\times10^{-05}$ & -.107651403$\times10^{ 01}$ &  .227867903$\times10^{ 04}$ & -.250497022$\times10^{ 07}$ &  .130409507$\times10^{ 10}$ & -.306635357$\times10^{ 12}$ &  .324675891$\times10^{ 13}$ &  .117450359$\times10^{ 17}$ & -.153259798$\times10^{ 19}$ \\
{\bf 14~$\to$~ 4}   &   .932853710$\times10^{-03}$ & -.147201544$\times10^{ 01}$ &  .878564114$\times10^{ 04}$ & -.171341106$\times10^{ 08}$ &  .160118156$\times10^{ 11}$ & -.836155976$\times10^{ 13}$ &  .250280121$\times10^{ 16}$ & -.402109207$\times10^{ 18}$ &  .269159361$\times10^{ 20}$ \\
{\bf 14~$\to$~ 5}   &   .529868835$\times10^{-04}$ & -.116360774$\times10^{ 01}$ &  .853351599$\times10^{ 04}$ & -.237273211$\times10^{ 08}$ &  .297376803$\times10^{ 11}$ & -.198853414$\times10^{ 14}$ &  .734439404$\times10^{ 16}$ & -.141300588$\times10^{ 19}$ &  .110533674$\times10^{ 21}$ \\
\end{longtable}
\normalsize
\renewcommand{\thefootnote}{\arabic{footnote}}
\renewcommand{\arraystretch}{1.0}
\end{landscape}

\newpage

\begin{landscape}
\renewcommand{\thefootnote}{\fnsymbol{footnote}}
\renewcommand{\arraystretch}{0.81}
\scriptsize
\setlength\tabcolsep{1.2ex}
\begin{longtable}{lccccccccc}
\caption[table10]{List of the parameters used in Eq.(\ref{eqn:BeH_DR_Interpolation}) for the VE and VdE Maxwell rate coefficients of BeH$^+$  ($v^+_i$~=~16 and 17) represented in the upper panels of Graph~\ref{fig:7}.}\label{tab:BeH_VE_Interpolation16-17}\vspace{0.5cm}\\
\hline \\[-2ex]
   \multicolumn{1}{c}{$v^+_i \to v^+_f$} &
   \multicolumn{1}{c}{$A_{{v^+_i}\to {v^+_f}}$} &
   \multicolumn{1}{c}{$\alpha_{{v^+_i}\to {v^+_f}}$} &
   \multicolumn{1}{c}{$B_{{v^+_i}\to {v^+_f}}(1)$} &
   \multicolumn{1}{c}{$B_{{v^+_i}\to {v^+_f}}(2)$} &
   \multicolumn{1}{c}{$B_{{v^+_i}\to {v^+_f}}(3)$} &
   \multicolumn{1}{c}{$B_{{v^+_i}\to {v^+_f}}(4)$}  &
   \multicolumn{1}{c}{$B_{{v^+_i}\to {v^+_f}}(5)$} &
   \multicolumn{1}{c}{$B_{{v^+_i}\to {v^+_f}}(6)$} &
   \multicolumn{1}{c}{$B_{{v^+_i}\to {v^+_f}}(7)$} \\[5pt]
 \hline \\[-2ex]
\endhead
{\bf 16~$\to$~ 0}   &   .760415137$\times10^{-07}$ & -.781552211$\times10^{ 00}$ &  .362270227$\times10^{ 04}$ & -.864034487$\times10^{ 07}$ &  .992100746$\times10^{ 10}$ & -.635687548$\times10^{ 13}$ &  .230529156$\times10^{ 16}$ & -.440949132$\times10^{ 18}$ &  .345064186$\times10^{ 20}$ \\
{\bf 16~$\to$~ 1}   &   .944911316$\times10^{-09}$ & -.277153578$\times10^{ 00}$ & -.122586357$\times10^{ 04}$ &  .222446721$\times10^{ 07}$ & -.205720792$\times10^{ 10}$ &  .105966910$\times10^{ 13}$ & -.307026813$\times10^{ 15}$ &  .466655684$\times10^{ 17}$ & -.288319559$\times10^{ 19}$ \\
{\bf 16~$\to$~ 2}   &   .147592850$\times10^{-06}$ & -.795476782$\times10^{ 00}$ & -.125105379$\times10^{ 04}$ &  .455817442$\times10^{ 07}$ & -.612779263$\times10^{ 10}$ &  .426089685$\times10^{ 13}$ & -.161937603$\times10^{ 16}$ &  .318740427$\times10^{ 18}$ & -.254049000$\times10^{ 20}$ \\
{\bf 16~$\to$~ 3}   &   .224210589$\times10^{-05}$ & -.107397045$\times10^{ 01}$ &  .836517193$\times10^{ 03}$ &  .749348278$\times10^{ 06}$ & -.225072475$\times10^{ 10}$ &  .191470873$\times10^{ 13}$ & -.793679766$\times10^{ 15}$ &  .163566913$\times10^{ 18}$ & -.134008894$\times10^{ 20}$ \\
{\bf 16~$\to$~ 4}   &   .152753881$\times10^{-04}$ & -.120048163$\times10^{ 01}$ &  .609033767$\times10^{ 04}$ & -.101974745$\times10^{ 08}$ &  .862664644$\times10^{ 10}$ & -.425592934$\times10^{ 13}$ &  .124042990$\times10^{ 16}$ & -.198132691$\times10^{ 18}$ &  .133707182$\times10^{ 20}$ \\
{\bf 16~$\to$~ 5}   &   .238270575$\times10^{-03}$ & -.144261430$\times10^{ 01}$ &  .964504824$\times10^{ 04}$ & -.244726088$\times10^{ 08}$ &  .287529593$\times10^{ 11}$ & -.182800059$\times10^{ 14}$ &  .648650802$\times10^{ 16}$ & -.120864113$\times10^{ 19}$ &  .921340693$\times10^{ 20}$ \\
\end{longtable}
\normalsize
\renewcommand{\thefootnote}{\arabic{footnote}}
\renewcommand{\arraystretch}{1.0}
\end{landscape}

\end{document}